\documentclass[journal]{IEEEtran}

\usepackage{caption}
\usepackage{algpseudocode}
\usepackage{amsmath}
\usepackage{amsfonts}
\usepackage{amssymb}
\usepackage{subfigure}
\usepackage{csquotes}
\usepackage{graphicx}
\usepackage{float}
\usepackage{wasysym}
 \usepackage{url}
 \usepackage{upgreek}
 \usepackage{wrapfig}
\usepackage{url}

\makeatletter
\def\url@leostyle{%
  \@ifundefined{selectfont}{\def\UrlFont{\sf}}{\def\UrlFont{\small\ttfamily}}}
\makeatother

\newcommand{\eat}[1]{}
\usepackage[linesnumbered,ruled]{algorithm2e}
\usepackage{fancybox}
\usepackage{eqnarray}
\usepackage{wrapfig}
\usepackage{enumerate}
\usepackage{booktabs}  
\usepackage{multirow}
\usepackage{threeparttable}

\usepackage{enumitem}

\usepackage{mdframed} 
\usepackage{arydshln}
\newenvironment{packed_enum}{%
  \begin{enumerate}%
  }{\end{enumerate}}

\usepackage{amsthm}
\newtheorem{theorem}{Theorem}

\newtheorem{definition}{\textbf{\textsc{Definition}}}

\usepackage{color}
\usepackage{ulem}

\begin{document}

\title{Privacy in Internet of Things: \\from Principles to Technologies}

\author{Chao~Li,~\IEEEmembership{Student Member,~IEEE,}
and Balaji~Palanisamy,~\IEEEmembership{Member,~IEEE}
        
\IEEEcompsocitemizethanks{\IEEEcompsocthanksitem Chao Li and Balaji Palanisamy are with the School of Computing and Information, University
of Pittsburgh.\protect\\
E-mail: chl205@pitt.edu, bpalan@pitt.edu}}

\maketitle

\begin{abstract}
Ubiquitous deployment of low-cost smart devices and widespread use of high-speed wireless networks have led to the rapid development of the Internet of Things (IoT). IoT embraces countless physical objects that have not been involved in the traditional Internet and enables their interaction and cooperation to provide a wide range of IoT applications. 
Many services in the IoT may require a comprehensive understanding and analysis of data collected through a large number of physical devices that challenges both personal information privacy and the development of IoT.
Information privacy in IoT is a broad and complex concept as its understanding and perception differ among individuals and its enforcement requires efforts from both legislation as well as technologies. 
In this paper, we review the state-of-the-art principles of privacy laws, the architectures for IoT and the representative privacy enhancing technologies (PETs). We analyze how legal principles can be supported through a careful implementation of privacy enhancing technologies (PETs) at various layers of a layered IoT architecture model to meet the privacy requirements of the individuals interacting with IoT systems. We demonstrate how privacy legislation maps to privacy principles which in turn drives the design of necessary privacy enhancing technologies to be employed in the IoT architecture stack.

\end{abstract}


\section{Introduction}
\label{s1}
\IEEEPARstart{U}{biquitous} deployment of low-cost smart devices and widespread use of high-speed wireless networks have led to the rapid development of Internet of Things (IoT). 
IoT embraces countless physical objects embedded with Radio Frequency Identification (RFID) tags, sensors and actuators that have not been involved in the traditional Internet and enables their interaction and cooperation through both traditional as well as IoT-specific communication protocols~\cite{Atzori2010,Malina2016}. 
Gartner~\cite{Gartner2017} estimates that around 20.4 billion `things' will be connected by the year 2020. These pervasive and heterogeneous devices that interact with the physical and digital worlds have the potential to significantly enhance the quality of life for individuals interacting with the IoT. With smart home and wearable devices, users obtain seamless and customized services from digital housekeepers, doctors and fitness instructors~\cite{Weinberg2015}. 
Smart building and smart city applications provide an increased awareness of the surroundings and offer greater convenience and benefits to the users~\cite{ENISA2015,Perera2014}.

Many services offered by IoT may require a comprehensive understanding of user interests and preferences, behavior patterns and thinking models. For instance, in the Christmas special episode of the British series `Black Mirror', the soul of a woman is copied to serve as the controller of her smart home, which can wake up the woman with her favorite music and breakfast as the copy knows her as no one else can~\cite{BlackMirror-netflix}. Such a digital copy, which could be hard to create in the traditional Internet, is relatively easier to be generated in the IoT era. While some individuals prefer the convenience of the services, some others may be concerned about their personal data being shared~\cite{Solove2006}. 
In 2013, the IEEE Internet of Things survey showed that 46\% of respondents consider privacy concerns as the biggest challenge for IoT adoption~\cite{IEEE2013}. Large scale data collection in the IoT poses significant privacy challenges and may hamper the further development and adoption by privacy-conscious individuals~\cite{Ziegeldorf2014}.

Information privacy is a broad and complex notion as its understanding and perception differ among individuals and its enforcement requires efforts from both legislation and technologies~\cite{ENISA2015,Cranor2012}.
Privacy laws help to enforce compliance and accountability of privacy protection and make privacy protection a necessity for every service provider~\cite{Cranor2012}.
Privacy enhancing technologies (PETs) on the other hand support the underlying principles guided by privacy laws that enable privacy protection strategies to be implemented in engineering~\cite{ENISA2015-2,Hoepman2014}.
In this paper, we study the privacy protection problem in IoT through a comprehensive review by jointly considering three key dimensions, namely the state-of-the-art principles of privacy laws, architectures for the IoT system and representative privacy enhancing technologies (PETs). Based on an extensive analysis along these three dimensions, we show that IoT privacy protection requires significant support from both privacy enhancing technologies (PETs) and their enforcement through privacy legislation.
We analyze how legal principles can be supported through a careful implementation of various privacy enhancing technologies (PETs) at various layers of a layered IoT architecture model to meet the privacy requirements of the individuals interacting with IoT systems. Our study is focused on providing a broader understanding of the state-of-the-art principles in privacy legislation associated with the design of relevant privacy enhancing technologies (PETs) and on demonstrating how privacy legislation maps to privacy principles which in turn drives the design of necessary privacy enhancing technologies to be employed in the IoT architecture stack.

We organize the paper in the following manner. In Section~\ref{section2}, we analyze the principles of privacy laws and present the privacy-by-design strategies that can adopt the general principles to engineering practice. 
In Section~\ref{section3}, we introduce the IoT system using a layered reference architecture and describe the functionalities and enabling technologies of each layer. We discuss how privacy-by-design strategies can be integrated into the reference architecture. 
In Section~\ref{section4} to Section~\ref{section6}, we introduce the state-of-the-art privacy enhancing technologies (PETs), analyze their suitability for privacy-by-design strategies and discuss the pros and cons of their use and implementation in each IoT layer.
In Section~\ref{s7}, we discuss privacy issues in IoT applications.
Finally, we present the related work in Section \ref{s8} and conclude in Section \ref{s9}.

\section{Privacy}
\label{section2}
Privacy is a complex and a subjective notion as its understanding and perception differ among individuals. In this section, we review the definitions of privacy in the past, introduce the privacy laws and analyze the state-of-the-art privacy legislation. We then introduce the privacy-by-design (PbD) strategies that facilitate the design of privacy-preserving systems satisfying the legal principles.

\subsection{Definition}
As far back as the thirteenth century, when the eavesdroppers were claimed to be guilty, the notion of media privacy had come into being~\cite{Langheinrich2001}. 
Then, with the technical and social development, the notion of privacy successively shifted to territorial (eighteenth century), communication (1930s), and bodily privacy (1940s)~\cite{Cranor2012}. Finally, in the 1960s, it was the rise of electronic data processing that brought into being the notion of information privacy (or data privacy) that has achieved lasting prominence until now. 
In 1890, Warren and Brandeis defined privacy as `the right to be let alone' in their famous article `The Right to Privacy'~\cite{Warren1890}. 
After that, many privacy definitions have been emerging unceasingly, but the one proposed by Alan Westin in his book `Privacy and Freedom' has become the base of several modern data privacy principles and law~\cite{Cranor2012}. Westin defined privacy as `the claim of individuals, groups, or institutions to determine for themselves when, how, and to what extent information about them is communicated to others'~\cite{Westin1968}, which mainly emphasized the control of the data subjects over their data. The authors in~\cite{Ziegeldorf2014} argued that Westin's definition was too general for the IoT area and they proposed a more focused one that defines the IoT privacy as the threefold guarantee including `awareness of privacy risks imposed by smart things and services surrounding the data subject; individual control over the collection and processing of personal information by the surrounding smart things; awareness and control of subsequent use and dissemination of personal information by those entities to any entity outside the subject’s personal control sphere'.

\subsection{Legislation}
Privacy laws form a critical foundation in the design of any privacy-preserving system. 
As the cornerstone of most modern privacy laws and policies, the Fair Information Practices (FIPs) are a set of internationally recognized practices to protect individual information privacy~\cite{Gellman2016}. The code of FIPs was born out of a report from the Department of Health, Education \& Welfare (HEW)~\cite{HEW1973} in 1973 and then adopted by the US Privacy Act of 1974, the most famous privacy legislation in the early stage. The original HEW FIPs consist of five principles that can be summarized as~\cite{HEW1973}:

\begin{mdframed}[innerleftmargin=1.6pt]
\begin{packed_enum}
  \item  No secret systems of personal data.
  \item  Ability for individuals to find out what is in the record, and how it is used.
  \item  Ability for individuals to prevent secondary use.
  \item  Ability to correct or amend records.
  \item  Data must be secure from misuse.
\end{packed_enum}
\end{mdframed}

\noindent However, as a federal law, the US Privacy Act of 1974 only works with the federal government. There is no general information privacy legislation that covers all states and areas~\cite{Levin2005}. As a result, the FIPs always act as the guideline of the various privacy laws and regulations ranging from different organizations (e.g., Stanford University~\cite{stanford}, Department of Homeland Security~\cite{DHS}) to different areas (e.g., HIPAA~\cite{HIPAA}, COPPA~\cite{COPPA}). 

In 1980, based on the core HEW FIPs, the Organization for Economic Cooperation and Development (OECD) adopted the Guidelines on the Protection of Personal Privacy and
Transborder Flows of Personal Data~\cite{OECD}. It is considered a historical milestone as it represented the first internationally-agreed upon privacy protection~\cite{Levin2005}. The eight principles extended from the five basic FIPs have been the foundation of most EU privacy laws later. 
They can be summarized as:

\begin{mdframed}[innerleftmargin=1.6pt]
\begin{packed_enum}
  \item  \textbf{Collection Limitation}: Collection should be lawful, fair and with knowledge or consent of the data subject.
  \item  \textbf{Data Quality}: Personal data should be purpose-relevant, accurate, complete and kept up-to-date.
  \item  \textbf{Purpose Specification}: Purposes should be specified earlier than collection and complied with.
  \item  \textbf{Use Limitation}: Personal data should not be disclosed, made available or used for non-specified purposes.
  \item  \textbf{Security Safeguards}: Personal data should be protected by reasonable security safeguards.
  \item  \textbf{Openness}: There should be a general policy of openness about developments, practices and policies with respect to personal data.
  \item  \textbf{Individual Participation}: An individual should have the right to access his data, be timely informed on data collection, be given disputable reason for denied lawful request and challenge his data to have the data erased, rectified, completed or amended.
  \item  \textbf{Accountability}: A data controller should be accountable for complying with measures which give effect to the principles stated above. 
\end{packed_enum}
\end{mdframed}

\noindent Although the OECD guidelines achieved worldwide recognition, it was nonbinding. It was not until 1995 that the EU passed Directive 95/46/EC~\cite{Directive1995} and the OECD guidelines were incorporated into an influential privacy law for the first time. Unlike the US, the EU dedicated to enforcing the omnibus privacy laws to comprehensively protect individual data in its member countries through not only the principles, but the restriction on the data transference with non-EU countries, which in turn has influenced the development of the privacy laws in the non-EU countries and the appearance of the data exchange regulations such as the Safe Harbor~\cite{safeHarbor} and its later replacement, the EU-US Privacy Shield framework~\cite{privacyShield}.
Recently, as the successor of Directive 95/46/EC, the General Data Protection Regulation (GDPR)~\cite{GDPR} was adopted by the EU in 2016 and it has come into force in 2018. In the GDPR, most principles are covered by the Article 5, including `lawfulness, fairness and transparency', `purpose limitation', `data minimization', `accuracy', `storage limitation', `integrity and confidentiality' and `accountability'. 
Its key changes in terms of the principles, compared with the Directive 95/46/EC, include six aspects~\cite{GDPR,GDPR2}:

\begin{itemize}[leftmargin=*]
\item \textbf{Consent}: The GDPR is more strict with consents. A consent should be graspable, distinguishable and easy to be withdrawn. 
\item \textbf{Breach Notification}: The GDPR makes the breach notification mandatory. The notification should be sent within 72 hours after being aware of the breach.
\item \textbf{Right to Access}: The first right mentioned in the OECD `individual participation' principle is strengthened in the GDPR.
\item \textbf{Right to be Forgotten}:  In the Article 17, data is required to be erased when the personal data are no longer necessary in relation to the purposes or the consent is withdrawn.
\item \textbf{Data Portability}: In the Article 20, a data subject has the right to receive his uploaded data in a machine-readable format and transmit it to another data controller.
\item \textbf{Privacy by Design}: The privacy by design is finally integrated into the privacy legal framework. As claimed in the Article 25, `the controller shall, both at the time of the determination of the means for processing and at the time of the processing itself, implement appropriate technical and organizational measures', which asks the privacy to be taken into account at the design stage, rather than as an add-on function.
\end{itemize}

\begin{figure*}
\centering
{
   
    \includegraphics[width=13cm,height=6.5cm]{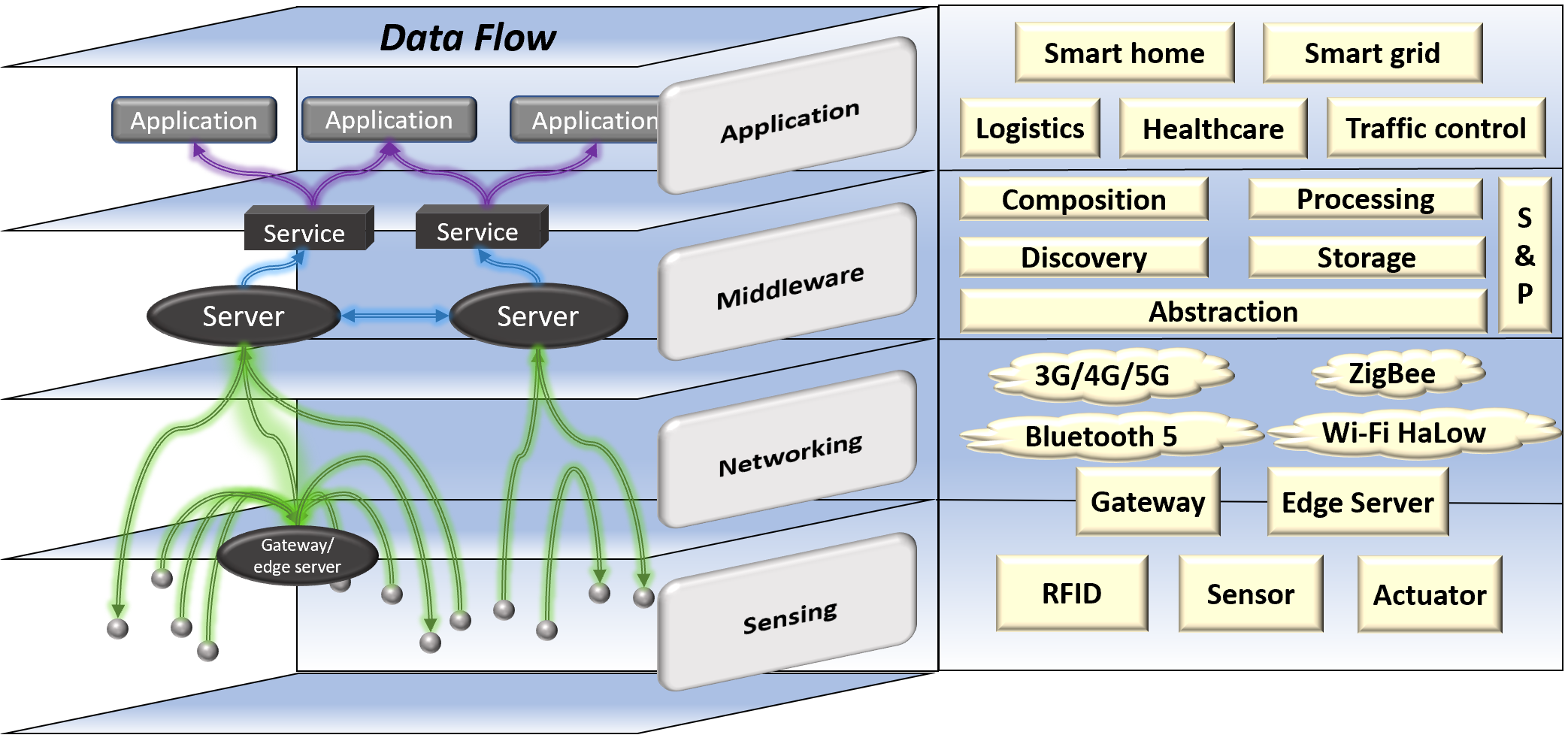}
}
\caption {IoT architecture and data flow~\cite{ETSI2013,Bonomi2012,CASAGRAS,ETSI}}
\label{s2} 
\end{figure*}

\subsection{Privacy by design} 
The notion of privacy by design (PbD)~\cite{ENISA2015,Hoepman2014,Langheinrich2001}, namely embedding the privacy measures and privacy enhancing technologies (PETs) directly into the design of software or system, is not new. As early as 2001, Langheinrich~\cite{Langheinrich2001} proposed six principles to guide the PbD in the ubiquitous systems, including \textit{notice}, \textit{choice and consent}, \textit{proximity and locality}, \textit{anonymity and pseudonymity}, \textit{security}, and \textit{access and recourse}. However, the PbD has never been extensively used in engineering. The main reason for its rare adoption is that most engineers either neglect the importance of privacy or refuse their responsibility on it~\cite{Spiekermann2009,Lahlou2005}. 

A privacy law may also need support from technologies.
In IoT, the enforcement of each principle in a privacy law may need to be supported by a set of technologies (e.g., PETs) in one or multiple layers. 
Here, the principles in the laws are usually described with very general and broad terms~\cite{ENISA2015} that makes it hard for engineers to properly implement them in the system design. 
Also, the availability of so many technologies makes the engineers' job of mapping technologies to principles difficult.
Therefore, we need the PbD to take the role as an adaptation layer between laws and technologies to translate legal principles to more engineer-friendly principles that can facilitate the system design.

In~\cite{Spiekermann2009}, Spiekermann and Cranor divided the technologies into two types of approaches to enable privacy in engineering, namely `\textit{privacy-by-architecture}' and `\textit{privacy-by-policy}'.
The privacy-by-architecture approaches can protect higher-level privacy through technologies enabling data minimization and local processing but is rarely adopted because of the lack of legal enforcement at that time and its conflict with the business interests. In contrast, the privacy-by-policy approaches protect only the bottom-line privacy through technologies supporting the notice and choice principles when the privacy-by-architecture technologies are not implemented. The authors argued that the privacy-by-policy technologies become less important when rigorous minimization has been guaranteed by the privacy-by-architecture technologies. 
Based on the two approaches, in 2014, Hoepman~\cite{Hoepman2014} proposed eight privacy design strategies, including four data-oriented strategies and four process-oriented strategies that roughly match the privacy-by-architecture and privacy-by-policy classification~\cite{ENISA2015-2,Hoepman2014}: 

\begin{mdframed}[innerleftmargin=1.6pt]
\textbf{Data-oriented strategies:}
\begin{packed_enum}
  \item  \textbf{Minimize}: The amount of processed personal data should be restricted to the minimal amount possible.
  \item  \textbf{Hide}: Any personal data, and their interrelationships, should be hidden from plain view.
  \item  \textbf{Separate}: Personal data should be processed in a distributed fashion, in separate compartments whenever possible.
  \item  \textbf{Aggregate}: Personal data should be processed at the highest level of aggregation and with the least possible
detail in which it is (still) useful.
\end{packed_enum}
\textbf{Process-oriented strategies:}
\begin{packed_enum}
  \item  \textbf{Inform}: Data subjects should be adequately informed whenever personal data is processed.
  \item  \textbf{Control}: Data subjects should be provided agency over the processing of their personal data.
  \item  \textbf{Enforce}: A privacy policy compatible with legal requirements should be in place and should be enforced.
  \item  \textbf{Demonstrate}: Be able to demonstrate compliance with the privacy policy and any applicable legal requirements.
\end{packed_enum}
\end{mdframed}

\noindent These strategies proposed by Hoepman not only inherit and develop the two engineering privacy approaches proposed by Spiekermann and Cranor, but also support the legal principles and PbD enforcement of the GDPR~\cite{Hoepman2014}. As a good combination point between legal principles and privacy enhancing technologies(PETs), these privacy design strategies have been widely accepted by recent work on privacy to fill the gap between legislation and engineering~\cite{ENISA2015,ENISA2015-2,Koops2014}. Therefore, we also adopt the eight privacy design strategies in this paper and study their relevant IoT layers (Section~\ref{section34}) and enabling PETs (Section~\ref{section4} to Section~\ref{section6}) in the context of IoT privacy.

\section{Privacy Protection in a Layered IoT Architecture}
\label{section3}
Several existing Internet-of-Things systems are designed using a layered architecture~\cite{ETSI2013,Bonomi2012,CASAGRAS,ETSI}.
In an IoT system, data is usually collected by end devices, transmitted through communication networks, processed by local/remote servers and finally provided to various applications. Thus, private data as it flows through multiple layers of the architecture stack, needs privacy protection at all layers. 
Here, implementing proper privacy design strategies based on the roles of the layers in the lifecycle of the data is important. 
Otherwise, techniques implemented at a specific layer may become either insufficient (privacy is breached at other layers) or redundant (privacy has been protected by techniques implemented at other layers).  
In this section, we introduce the reference IoT architecture adopted in this study and present the IoT privacy protection framework that shows how to integrate the privacy design strategies in the layered IoT architecture.

\subsection{Reference IoT architecture}

\begin{table}[tp]  
  \centering   
  \caption{IoT architecture comparison ($\surd$ contained as a separate layer, \Circle\ merged with other layers, $\times$ not contained)}  
  \label{tab:33} 
\begin{tabular} 
%{cccccc}  
{p{2.3cm}<{\centering} p{1.0cm}<{\centering} p{1.1cm}<{\centering} p{1.2cm}<{\centering} p{1.1cm}<{\centering}}
\hline
\hline 
{\bf Source} & {\bf Perception} & {{\bf Networking}} & { {\bf Middleware}} & {\bf Application}\\ 
\hline
IEEE~\cite{Minerva2015} & $\surd$  & $\surd$  & $\times$ & $\surd$\\
M2M~\cite{ETSI} & $\surd$  & \Circle  & \Circle & \Circle\\
oneM2M~\cite{ETSI2013} & $\times$  & $\surd$  & $\surd$ & $\surd$\\
CASAGRAS~\cite{CASAGRAS} & $\surd$  & $\surd$  & \Circle & \Circle\\
Cisco~\cite{Bonomi2012} & $\surd$  & $\surd$  & \Circle & \Circle\\
Soma~\textit{et cl.}~\cite{bandyopadhyay2011role} & $\surd$  & $\times$  & $\surd$ & $\surd$\\
Addo~\textit{et cl.}~\cite{addo2014reference} & $\surd$  & $\surd$  & $\surd$ & $\times$\\
Funke~\textit{et cl.}~\cite{Funke2015} & $\surd$  & $\surd$  & $\surd$ & $\times$\\
Sun~\textit{et cl.}~\cite{Sun2014} & $\surd$  & $\surd$  & \Circle & \Circle\\
Perera~\textit{et cl.}~\cite{Perera2016} & $\surd$  & $\surd$  & \Circle & \Circle\\
Dabbagh~\textit{et cl.}~\cite{dabbagh2017internet} & $\surd$  & $\surd$  & $\surd$ & $\surd$\\
Botta~\textit{et cl.}~\cite{botta2016integration} & $\surd$  & $\times$  & $\surd$ & $\surd$\\
\hline
\hline 
\end{tabular}
\end{table}

In general, the number of layers proposed for the architecture of IoT varies considerably. 
After reviewing a number of existing IoT architectures, we adopt a four-layer architecture as the reference IoT architecture in this paper, which consists of perception layer, networking layer, middleware layer and application layer.
The adoption of the four-layer reference architecture in our study has two key benefits. 
First, the importance of each layer in the four-layer architecture has been recognized by most existing architectures as the four-layer architecture allows a comprehensive view of privacy in IoT.
As shown in Table~\ref{tab:33}, several existing architectures~\cite{ETSI,CASAGRAS,Bonomi2012,Sun2014,Perera2016,dabbagh2017internet} include all the four layers, either as separate layers or integrated layers. Second, as the four-layer architecture is the most fine-grained model among all the candidate architectures, it allows a detailed and fine-grained analysis of privacy protection at different layers and avoids possible lack of differentiation when the layers are not distinct as in~\cite{ETSI,CASAGRAS,Bonomi2012,Sun2014,Perera2016}.

\begin{figure*}
\centering
\subfigure[{\small Active collection}]
{
   \label{f42}
   \includegraphics[width=7cm,height=3.5cm]{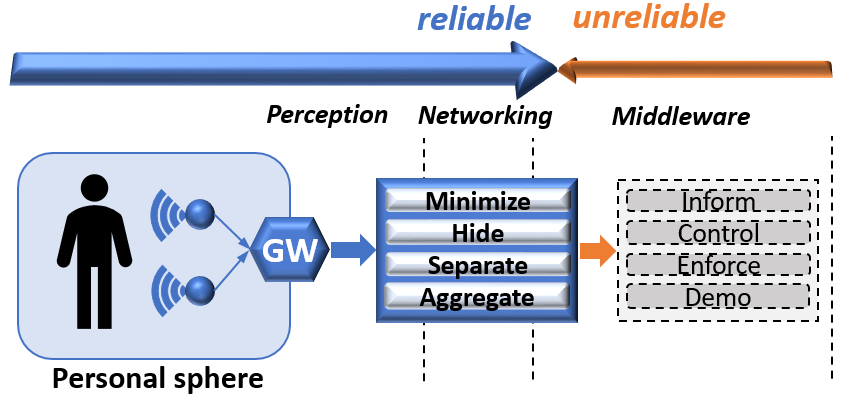}
}
\subfigure[{\small Passive collection}]
{
	\label{f43}
    \includegraphics[width=7cm,height=3.5cm]{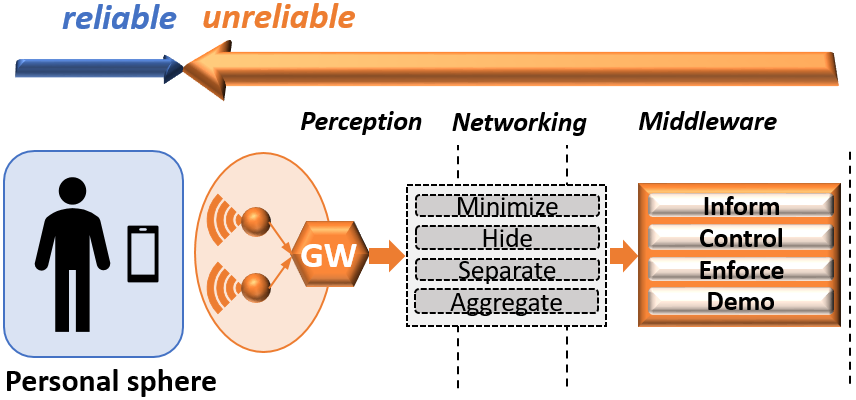}
}
\caption{IoT privacy protection framework~\cite{Ziegeldorf2014,Hoepman2014,Spiekermann2009,Henze2016}}
\label {r4}
\end{figure*}

As the lowest layer of the architecture (Fig.~\ref{s2}), perception layer works as the base of entire Internet of Things. It bridges the gap between physical world and digital world by making innumerable physical entities identifiable (e.g., RFIDs~\cite{Jia2012}), perceptible (e.g., sensors~\cite{Atzori2010}) and controllable (e.g., actuators~\cite{Gubbi2013}) to enable deep interaction between physical and digital worlds~\cite{Minerva2015,Tan2010}.
The networking layer plays a pivotal role to link the perception layer and middleware layer so that sensed data and corresponding commands can be seamlessly transmitted between the two layers.
Unlike the traditional Internet, the vast number of heterogeneous power-limited devices in the perception layer and the various application scenarios in the application layer create a vital need for communication technologies that support low energy consumption, low latency, high data rate and high capacity. Main techniques supporting IoT networking layer include ZigBee~\cite{specification2006zigbee}, Bluetooth 5~\cite{Bluetooth}, Wi-Fi HaLow~\cite{Wi-Fi} and 5th generation mobile networks~\cite{ITU-R}.
The middleware layer works as the `brain' of IoT to process the numerous data received from lower layers. 
To cope with the interoperability of the heterogeneous physical devices~\cite{Bandyopadhyay2011,Chaqfeh2012}, the \textit{device abstraction} component semantically describes the resources with a consistent language such as the eXtensible Markup Language (XML), Resource Description Framework (RDF) or Web Ontology Language (OWL)~\cite{aberer2006middleware,Eisenhauer2010,Gómez-Goiri2010}. 
Based on that, resources are made discoverable through the \textit{resource discovery} component by using Semantic Annotations for WSDL and XML Schema (SAWSDL)~\cite{Eisenhauer2010} or simply key words~\cite{Node-Red}. 
Then, if needed, multiple resources can be composed through the \textit{composition} component~\cite{Atzori2010,Ngu2017} to enhance their functionality.
After that, received data could be stored (\textit{storage} component) in either cloud or databases and kept available to be queried.
Different computational and analytical units can be combined to form the \textit{processing} component. 
Here, security of data, namely their confidentiality, integrity, availability, and non-repudiation~\cite{Fremantle2015} need to be well protected. If data can make its owner either identified or identifiable, privacy enhancement technologies (PETs) are necessary to protect privacy so that privacy principles required by the laws can be satisfied.
As the highest layer of the architecture, the application layer contains various IoT applications that have been widely studied in past literature~\cite{Atzori2010,Weinberg2015}. Depending on the scenarios that the private data is collected, different applications may encounter different privacy issues.

\subsection{IoT privacy protection framework}
\label{section34}
In this section, we study the integration of privacy design strategies in the layered architecture by introducing an IoT privacy protection framework. 
From the viewpoint of the IoT architecture, data is collected by devices at perception layer and transmitted to middleware layer through networking layer, which makes data move away from the control of data subjects~\cite{Solove2006,Ziegeldorf2014}.
We can apply the notion of personal sphere~\cite{Ziegeldorf2014,Henze2016} to assist interpretation. A personal sphere consists of a set of personal devices and a gateway, both trusted by data subjects. In some cases, gateway can be assisted by a more powerful trusted third party (TTP).
Data collected by these personal devices has to be passed to the gateway and/or the TTP to be processed before being transmitted to the data controllers that data subjects distrust. Such a personal sphere is quite important to implement the four data-oriented privacy design strategies because it offers a reliable platform to make the raw data minimized, hidden, separated and aggregated. As pointed out by Spiekermann and Cranor~\cite{Spiekermann2009}, once sensitive information in data has been adequately constrained through PETs such as homomorphic encryption~\cite{Gentry2009} and $k$-anonymization~\cite{Sweeney2002}, the privacy-by-policy approaches, namely the four process-oriented strategies become less important. In IoT, such a personal sphere can be created when data is actively collected, as shown in Fig.~\ref{f42}. For example, smart appliances and home router form an indoor personal sphere while wearable devices and smartphones compose an outdoor personal sphere. The trusted local gateway and/or remote TTP is a critical element in the system, which allows data subjects to launch proper PETs to process data with the four data-oriented strategies.

Due to the invisibility of numerous IoT devices at perception layer, personal data may be sensed by untrusted devices outside the personal sphere and the data subjects may be completely unaware of the collection~\cite{Minerva2015,addo2014reference,ramos2014towards}. Such a passive collection makes data subjects lose control over their personal data at the earliest stage and provides no trusted platform to implement the four data-oriented privacy design strategies, as shown in Fig.~\ref{f43}. It is therefore the four process-oriented strategies that can play a more important role by promoting the power of data subjects when raw data is obtained by data controllers~\cite{Hoepman2014,Spiekermann2009}. Specifically, the \textit{inform} strategy and \textit{control} strategy enhance the interaction between data subjects and their data while the \textit{enforce} strategy and \textit{demonstrate} strategy force data controllers to comply with privacy policy and further require the compliance to be verifiable. As it is the remote data controllers that should offer proper PETs to support the four process-oriented strategies, these system-level strategies are primarily implemented at middleware layer, with the assistance of other layers. It is worth mentioning that we do not mean active collection only needs data-oriented strategies and passive collection only requires process-oriented strategies. In both cases, all the strategies are required to jointly work to support the legal principles. For example, although the single \textit{minimize} strategy is hard to be fulfilled in the passive collection, its implementation can be enforced and verified by process-oriented strategies.

In the next three sections, we present and evaluate PETs implemented at the perception layer, networking layer and middleware layer respectively.

\section{Privacy at perception Layer}
\label{section4}
We evaluate and compare the anonymization-based PETs and perturbation-based PETs that help to implement the \textit{Minimize} and \textit{Aggregate} strategies and present the encryption-based PETs that implements the \textit{Hide} strategy. These PETs are primarily implemented at the perception layer (e.g., local personal gateway, trusted edge server), but they can also be implemented at the middleware layer using a trusted third party (TTP). It is worth noting that the \textit{Separate} strategy is naturally achieved by local processing in the perception layer.

\subsection{Anonymization and perturbation}
\label{section42}

Both anonymization~\cite{Sweeney2000} and perturbation~\cite{dwork} techniques can fulfill the \textit{Minimize} and \textit{Aggregate} strategies by reducing the released information and increasing the granularity. The main difference between them is that the results of the anonymization are generalized while the results of the perturbation are with noises. 
In this section, we evaluate the representative anonymization and perturbation privacy paradigms, namely the $k$-anonymity~\cite{Sweeney2002} and differential privacy~\cite{Dwork2006} respectively, in terms of their practicability in IoT by analyzing their performance under the following IoT specific challenges~\cite{ENISA2015}:

\begin{itemize}[leftmargin=*]
\item Large data volume: The gateways may control thousands of sensors that collect massive data.
\item Streaming data processing: In some real-time IoT applications (e.g., traffic control), data may be seamlessly collected to form streams. 
\item Lightweight computation: Since the gateways (e.g., router and phone) are still resource-constrained devices, algorithms are expected to have low complexity.
\item Decentralized computation: In the IoT applications such as smart grid, the personal data may be collected by untrusted entities. Decentralization data aggregation may be employed under such scenarios.
\item Composability: The privacy should still be guaranteed after the data uploaded to the middleware layer is combined with other data.
\item Personalization: For most personal service IoT applications, each customer has different privacy understanding and requirements and there is a natural need for personalized solutions.
\end{itemize}

\subsubsection{Anonymization}
The traditional privacy-persevering data publication (PPDP) schemes typically involve four entities, namely data subject, data curator, data user and data attacker~\cite{Yu2016}. The data curator collects data from data subjects, processes the collected data and releases the privacy-preserving dataset to the data users. 
Usually, the collected data related to a data subject can be classified into four categories, namely explicit identifiers (e.g., names, SSN), quasi-identifiers (e.g., age, gender), sensitive attributes and non-sensitive attributes~\cite{Sweeney2002}.
In IoT, unlike the traditional Internet that requires all the records to be typed in, the identifiers are usually input through RFID tags and cameras. For example, vehicles can be identified by E-ZPass~\cite{Rieback2006} through RFID and individuals can be identified through RFID-enabled smart cards in shopping malls~\cite{Chung2016}. The sensitive and non-sensitive attributes are usually collected by sensors.

As a candidate PPDP approach, anonymization aims to cut off the connection between each record and its corresponding data subject so that the sensitive attributes cannot be linked with specific individuals~\cite{Benjamin2010}. Obviously, the explicit identifiers should be removed before publication for the privacy purpose. However, in 2000, Sweeney found that 87\% of US citizens can be uniquely re-identified by combining three quasi-identifiers, namely [ZIP, gender, date of birth]~\cite{Sweeney2000}. This linking attack has motivated the researchers to devise stronger anonymization paradigms including $k$-anonymity~\cite{Sweeney2002}, $l$-diversity~\cite{Machanavajjhala2007} and $t$-closeness~\cite{Li2007}, where $k$-anonymity~\cite{Sweeney2002} requires each quasi-identifier group to appear at least $k$ times in the dataset. 
We next discuss the use of anonymization in the context of IoT:

\noindent \textbf{Large volume:} The performance of anonymization algorithms may be affected by the dimensions of both rows and columns in the table, so the anonymization scheme is expected to be scalable for datasets with millions of records and multi-dimensional attributes. For the former, spatial indexing has been proved to be a good solution to handle numerous records in a dataset~\cite{LeFevre2006,Iwuchukwu2007}. One attribute can be efficiently $k$-anonymized through $B^{+}$ tree indexing and the $R^{+}$ tree indexing can be implemented to effectively generate non-overlapping partitions for tables with 100 million records and nine attributes~\cite{Iwuchukwu2007}. However, as analyzed by~\cite{Aggarwal2005}, the $k$-anonymity algorithms may work well for tables with a small number of attributes (e.g., 10) but not the ones with a large number of attributes (e.g., 50). The increasing number of attributes makes the number of combinations of dimensions exponentially increased and results in unacceptable information loss. Therefore, how to enhance the utility of $k$-anonymized datasets with a large number of attributes is still an open issue for future research. 
An anonymization method for the sparse high-dimensional binary dataset with low information loss was proposed in~\cite{Ghinita2008}, but there were no effective schemes for non-binary datasets.

\noindent \textbf{Streaming data:} There have been several strategies to anonymize data streams~\cite{Cao2011,Li2008}. In CASTLE~\cite{Cao2011}, a set of clusters of tuples are maintained and each incoming tuple in a stream is grouped into a cluster and generalized to the same level of other tuples in the cluster. Each tuple maintains a delay constraint $\delta$ and must be sent out before the deadline to make the processing real-time. At the end of $\delta$, if the cluster containing that tuple has at least $k$ members, all the tuples within it can be released. Otherwise, a cluster satisfying the $k$ requirement can be generated through a merge and split technique for the tuple and the information loss during the process can be minimized. In SKY~\cite{Li2008}, a top-down specialization tree is maintained and each incoming tuple is mapped to one node in the tree based on its attributes. Each node can be a work node or a candidate node depending on whether there have been at least $k$ tuples generalized and output from it. If the incoming node is mapped to a work node, it can be directly generalized and released. Otherwise, it has to wait for other arriving tuples at the node during the time $\delta$ or be generalized and released through the parent node at the end of $\delta$.

\noindent \textbf{Lightweight:} It has been proved that the optimal $k$-anonymity aiming to anonymize a table with minimum suppressed cells is NP-hard even when the attribute values are ternary~\cite{Meyerson2004,Aggarwal2005-2,Park2007}. The complexity of approximate algorithms for k-Anonymity has been reduced from $O(k \log k)$~\cite{Meyerson2004} to $O(k)$~\cite{Aggarwal2005-2} and later to $O(\log k)$~\cite{Park2007}.

\noindent \textbf{Collaboration:} Anonymization techniques can be implemented in a distributed manner. That is, multiple entities can locally anonymize their own table to make the integrated table $k$-anonymous without revealing any additional information during the process.
Several SMC protocols have been proposed to solve this problem~\cite{Wang2005,Jiang2006}. In~\cite{Wang2005}, a top-down specialization scheme was proposed to support joint anonymization between two parties. Specifically, the two parties first generalize their local table to the root. Then, in each iteration, they find the local specialization maximizing the ratio between information gain and privacy loss (IGPL) over the local table. The party with a higher IGPL wins the game in this iteration, applies its local specialization over its local table and then instructs the grouping in the table of the other party. For the same objective, a scheme based on cryptography was proposed in~\cite{Jiang2006}. Each party locally generalizes the local table and then jointly determines whether the integrated table is $k$-anonymous. If not, each party then generalizes its local table to the next layer and repeats the two steps. 

\noindent \textbf{Composability:} As shown in~\cite{ENISA2015}, the $k$-anonymity does not offer composability. That is, two $k$-anonymous datasets cannot guarantee their joint dataset is $k'$-anonymous ($k'>1$). Because of this, the integration of multiple $k$-anonymous datasets in the middleware layer can be a significant challenge.

\noindent \textbf{Personalization:} Most anonymization algorithms assume that all the record owners have same privacy preference. Therefore, less-anonymization can put privacy in risk but over-anonymization increases the information loss. 
To solve this, Xiao et al.~\cite{Xiao2006} organize the sensitive attributes in a top-down taxonomy tree and allow each record owner to indicate a guarding node. That is, the sensitive attribute of a specific record owner should be generalized to at least the guarding node in the taxonomy tree and the adversary has little opportunity to link the record owner with the child nodes of the guarding node that carry fine-grained information. Their algorithm first runs common $k$-anonymity algorithms over the quasi-identifiers and then generalizes the sensitive attribute through the taxonomy tree based on the claimed guarding nodes. Recently, Xu et al.~\cite{Xu2013} argued that the generalization of sensitive attributes results in information loss and they allow the record owners to claim the expected value of $k$. Their algorithm first achieves $k_{min}$-anonymity over the entire dataset, where $k_{min}$ is the minimum expected $k$ value, namely the most strict privacy requirement. Then, based on the data structure called d-dimensional quasi-attribute generalization lattice, some quasi-attributes can be merged to match the lower values of $k$ expected by some record owners.

\subsubsection{Differential privacy}

Differential privacy is a classical privacy definition~\cite{Dwork2006} that makes very conservative assumptions about the adversary's background knowledge and bounds the allowable error in a quantified manner. 
In general, differential privacy is designed to protect a single individual's privacy by considering adjacent data sets which differ only in one record. Before presenting the formal definition of $\epsilon$-differential privacy, we first define the notion of adjacent datasets in the context of differential privacy.
A data set $D$ can be considered as a subset of records from the universe $U$, represented by $D\in\mathbb{N}^{|U|}$, where $\mathbb{N}$ stands for the non-negative set and $D_i$ is the number of element $i$ in $\mathbb{N}$. For example, if $U=\{a,b,c\}$, $D=\{a,b,c\}$ can be represented as $\{1,1,1\}$ as it contains each element of $U$ once. Similarly, $D'=\{a,c\}$ can be represented as $\{1,0,1\}$ as it does not contain $b$. Based on this representation, it is appropriate to use $\mathit{l}_1$ distance (Manhattan distance) to measure the distance between data sets.

\begin{definition}[\textsc{Data set Distance}]
\textit{The $l_1$ distance between two data sets $D_1$ and $D_2$ is defined as $||D_1-D_2||_1$, which is calculated by:
$$||D_1-D_2||_1=\sum_{i=1}^{|U|}|D_{1_i}-D_{2_i}|$$
}
\end{definition}
The manhattan distance between the datasets leads us the notion of adjacent data sets as follows.
\begin{definition}[\textsc{Adjacent Data set}]
\textit{Two data sets $D_1$, $D_2$ are adjacent data sets of each other if $||D_1-D_2||_1=1$}.
\end{definition}

Based on the notion of adjacent datasets defined above, differential privacy can be defined formally as follows. In general, $\epsilon$-differential privacy is designed to protect the privacy between adjacent data sets which differ only in one record.

\begin{definition}[\textsc{Differential privacy}~\cite{Dwork2006}]
\label{dp}
\textit{A randomized algorithm $\mathcal{A}$ guarantees $\epsilon$-differential privacy if for all adjacent datasets $D_1$ and $D_2$ differing by at most one record, and for all possible results $\mathcal{S}\subseteq Range(\mathcal{A})$, $$Pr[\mathcal{A}(D_1)=\mathcal{S}]\leq e^{\epsilon}\times Pr[\mathcal{A}(D_2)=\mathcal{S}]$$ where the probability space is over the randomness of $\mathcal{A}$.}
\end{definition}

Many randomized algorithms have been proposed to guarantee differential privacy, such as the Laplace Mechanism\cite{Dwork2006}, the Gaussian Mechanism\cite{Dwork2013} and the Exponential Mechanism\cite{McSherry2007}. 
Given a data set $D$, a function $f$ and the budget $\epsilon$, the Laplace Mechanism first calculates the actual $f(D)$ and then perturbs this true answer by adding a noise\cite{Dwork2006}. The noise is calculated based on a Laplace random variable, with the variance $\lambda=\triangle f /\epsilon$, where $\triangle f$ is the $\mathit{l}_1$ sensitivity.
We next analyze differential privacy in terms of the challenges in the context of IoT:

\begin{table*}[tp]  
  \centering   
  \caption{Evaluation of $k$-anonymization and differential privacy (\Circle~Good \LEFTcircle~Not enough \CIRCLE~Poor)}  
  \label{tab:3} 
\begin{tabular} 
{ccccccc}  
\hline
\hline 
{\bf PET} & {\bf Large volume} & {{\bf Streaming data}} & { {\bf Lightweight}} & {\bf Collaboration} & {\bf Composability} & {\bf Personalization}\\ 
\hline
$k$-anonymization & \LEFTcircle  & \Circle  & \Circle & \Circle & \CIRCLE & \LEFTcircle \\
Differential privacy & \Circle  & \Circle & \LEFTcircle  & \Circle & \Circle & \LEFTcircle \\
\hline
\hline 
\end{tabular}
\end{table*}

\noindent \textbf{Large volume:} The large volume of data is naturally not a problem for differential privacy as the perturbation is usually implemented over the statistical value of the collected data.

\noindent \textbf{Streaming data:} There have been many works on applying differential privacy over streaming data since 2010~\cite{Dwork2010,Dwork2010-2}. The data stream was assumed to be a bitstream, where each bit can be either 1 or 0 representing if an event was happening or not at each timestamp. Mechanisms were proposed to protect either the event-level or user-level differential privacy, depending on whether a single event or all the events related to a single user can be hidden by the injected noise. The early works focused on event-level privacy. In~\cite{Dwork2010}, a counter was set to report the accumulated $1$s in the data stream at each timestamp and each update value can be added with a $Lap(\frac{1}{\epsilon})$ noise to guarantee the differential privacy. Furthermore, for a sparse stream with few $1$s, an update can be set to happen only after the number of $1$s has been accumulated over a threshold. Later in~\cite{Chan2011}, the noise error was reduced through using a binary tree data structure. Specifically, the nodes in the binary tree, except the leaf nodes, represent the sums of sections of consecutive bits in the stream and the Laplace noises were added to these nodes, instead of the leaf nodes. This scheme can effectively reduce the noise error from $O(T)$ to $O((\log T)^{1.5})$, where $T$ denotes the number of timestamps, namely the length of the stream. In~\cite{Fan2012}, the user-level privacy was supported and the noise error in this work was suppressed through sampling.

\noindent \textbf{Lightweight:} The complexity of differential privacy algorithms is quite variable on a case-by-case manner. If both the sensitivity and budget allocation are fixed, the complexity can be very low, as only one value is required to be sampled from a random distribution with fixed variance. However, in the cases that the sensitivity or budget allocation has to be calculated on the fly, the complexity will increase. 

\noindent \textbf{Collaboration:} Differential privacy for data aggregation is usually guaranteed by noises added through Laplace mechanism~\cite{Dwork2006}. A simple solution for this is to make the data aggregator directly aggregate the raw data received from data subjects and then add noise to it. However, in some scenarios such as smart metering, the aggregator (electricity supplier) may be untrusted~\cite{Ács2011} and may require the data subjects (smart meters) to locally add noise to perturb its raw data and then send the perturbed data to the aggregator so that the raw data is protected from the aggregator and the aggregated noise automatically satisfies the Laplace Mechanism. This distributed implementation of Laplace Mechanism, also known as Distributed Perturbation Laplace Algorithm (DLPA), has recently received attention from privacy researchers. The base of DLPA is the infinite divisibility feature of Laplace distribution~\cite{Kotz2012} that allows the noise sampled from Laplace distribution (central noise) to be the sum of $n$ other random variables (local noises). The local noise can still follow the Laplace distribution~\cite{Goryczka2013}. However, since a Laplace distributed random variable can be simulated by two gamma distributed random variables and four normal distributed random variables, the local noise can also follow the gamma distribution~\cite{Ács2011} or Gaussian distribution~\cite{Rastogi2010}. In~\cite{Goryczka2013}, the three schemes were compared and the Laplace distributed local noise was shown to be more efficient in terms of local noise generation.

\noindent \textbf{Composability:} Differential privacy offers strong composability:
\begin{theorem}[\textsc{Composition theorem}~\cite{Dwork2013}]
\textit{Let $\mathcal{A}_i$ be $\epsilon_i$-differential private algorithms applied to independent datasets $D_i$ for $i \in [1,k]$. Then their combination $\mathcal{A}_{\sum_{i=1}^k}$ is $max(\epsilon_i)$-differential private.}
\end{theorem}
In the middleware layer, multiple independent differentially private outputs can be combined and their integration still satisfies differential privacy. Differential privacy also satisfies the post-processing theorem, which further enhances its flexibility in the middleware layer.

\begin{theorem}[\textsc{Post-processing}~\cite{Dwork2013}]
\textit{Let $\mathcal{A}$ be a $\epsilon$-differentially private algorithm and $g$ be an arbitrary function. Then $g(\mathcal{A})$ is also $\epsilon$-differentially private.}
\end{theorem}

\noindent \textbf{Personalization:} In traditional differential privacy, the parameter $\epsilon$ is usually set globally for all the record owners. Recently, several works try to make it personalized. In~\cite{Jorgensen2015}, two solutions were proposed, based on sampling and Exponential Mechanism respectively. The first approach non-uniformly samples the records from the dataset with the inclusion probabilities related to the preferred privacy preferences (values of $\epsilon$). For each record, if the expected $\epsilon$ is smaller than a threshold $t$, it may only be selected with a probability related to the $\epsilon$. Otherwise, the record will be selected. Then, any $t$-differentially private mechanism can be applied to the sampled dataset. Their second approach is inspired by the Exponential Mechanism. Unlike the traditional Exponential Mechanism, to take personalization into account, the probability of each possible output values is computed based on the personalized privacy preferences (values of $\epsilon$).

\subsubsection{Anonymization vs. Differential privacy}
To sum up, as shown in Table \ref{tab:3}, both the techniques have similar features in terms of their support for streaming data, collaboration and personalization. Anonymization techniques are difficult to scale for datasets with many attributes while the complexity of differential privacy algorithms varies case by case. It is the composability feature that makes differential privacy a clear winner. Due to lack of composability, the operability and utility of the data protected by the $k$-anonymization paradigm are significantly constrained in the middleware layer.

\subsection{Encryption}
\label{section43}
Encryption techniques are not only the fundamental building block of security, but also the foundation of a large number of PETs in privacy. With respect to the eight privacy design strategies, encryption is the most direct supporter of the \textit{Hide} strategy, which also satisfies the `security safeguards' requirement of privacy laws. Therefore, in terms of IoT privacy, the role of encryption is twofold. On one hand, the commonly used cryptographic primitives, such as AES~\cite{NIST2001} and RSA~\cite{Rivest1983}, protect the security of every IoT layer so that the adversaries are prevented from easily compromising the confidentiality and integrity of data in IoT devices. From this perspective, the personal data is confined to a safe zone without being disclosed to unknown parties, thus also protecting the privacy of the data subject as the control over the data is enhanced. On the other hand, in IoT, the middleware may not be trusted or trustworthy but it is an indispensable stakeholder in most IoT applications. Hence, PETs such as homomorphic encryption~\cite{Gentry2009}, searchable encryption~\cite{Curtmola2011} and SMC~\cite{Yao1982} are required to make the middleware work without accessing the private information. 
Here, lightweight cryptography that can support encryption over devices with low capacity becomes a critical element in protecting IoT privacy. In this section, to comprehensively review the current state of work in this area, we first go through the real capacity of various types of IoT devices in the perception layer and evaluate the implementation of commonly used cryptographic primitives over them to see when and where lightweight cryptography is required. Then, we review the candidate lightweight solutions in each area of cryptography and present the NIST General Design Considerations~\cite{McKay2016}. Finally, we discuss the PETs aiming to blind the middleware and their performance over IoT devices. 

\noindent \textbf{The capacity of IoT devices}: The types of IoT edge devices in the perception layer range from  resource-rich devices such as computers and smartphones to resource-constrained devices such as embedded systems, RFID and sensors. For the resource-rich devices, the traditional cryptographic primitives work well for the encryption tasks. Thus, the lightweight cryptography techniques are mainly required by the resource-constrained devices that can not support traditional cryptographic primitives. This also requires the resource-rich devices in the middleware layer to adopt them in order to decrypt the data encrypted using lightweight cryptography techniques.
Most IoT embedded systems and intelligent systems are enabled by the 8-bit, 16-bit or 32-bit microcontrollers (MCUs) with highly restricted random-access memory (RAM) as low as 64 bytes RAM (e.g., NXP RS08KA, \$0.399)~\cite{NXP}. 
The RFID and sensor devices are usually more cost-sensitive and they employ the use of application specific integrated circuit (ASIC)~\cite{Saarinen2012}. Therefore, in hardware, the price of these devices is proportional to the area of ASIC in silicon, measured by the gate equivalents (GE), namely the ratio between the area of ASIC and the area of a two-input NAND gate~\cite{Rolfes2008}. The implementation of lightweight cryptography techniques over such devices has to meet several stringent conditions, including under 2000 GE to achieve low-cost, under 50 cycles for obtaining low-latency and less than $10\frac{\mu W}{MHz}$ average power usage for meeting low-energy requirements~\cite{Saarinen2012}.

\noindent \textbf{Traditional cryptographic primitives over constrained devices}: Most traditional commonly-used cryptographic primitives face severe challenges in the constrained environment. The AES-128~\cite{NIST2001} may be the most suitable lightweight block cipher because of its low number of rounds and small key size. In~\cite{Malina2016}, AES-128 was tested over several MCUs and smart cards and achieved $1.58ms$ execution time and 0.6kB RAM consumption over the MSP microcontrollers. The results show that AES works well for most MCUs, but not the ones with ultra-low RAM (e.g., NXP RS08KA). In terms of hash functions, the SHA-2 is acceptable to implement the cryptographic schemes requiring a few hash functions over the MSP microcontrollers with tens to hundreds of milliseconds execution time and 0.1kB RAM. However, as illustrated by Ideguchi et al.~\cite{Ideguchi2009}, the SHA-3 candidates cannot be supported by the low-cost 8-bit microcontrollers with 64 byte RAM. In the NIST competition, the lowest number of GE required by the SHA-3 is still 9200~\cite{Gauravaram2009}. Also, both the RSA~\cite{Rivest1983} for asymmetric encryption and elliptic curve point multiplication for ECDH and ECDSA schemes were found to be too high-cost for even the MSP microcontrollers~\cite{Malina2016}. 

\noindent \textbf{Attribute-Based Encryption in IoT}: Attribute-Based Encryption (ABE)~\cite{sahai2005fuzzy} is a promising mechanism to implement fine-grained access control over encrypted data. With ABE, an access policy can be enforced during data encryption, which only allows authorized users with the desired attributes (e.g., age, gender) to decrypt the data.
Depending on whether the access policy is associated with the key or ciphertext, Key-Policy ABE
(KP-ABE)~\cite{goyal2006attribute} and Ciphertext-Policy ABE (CP-ABE)~\cite{bethencourt2007ciphertext} were proposed, respectively.
Although ABE looks like the desired approach to secure data communication and storage in IoT with flexible access control, its implementation in IoT may encounter three main challenges.
First, 
current IoT applications only need IoT devices to encrypt data using public keys and hence, key management may not be a significant issue.
However, future autonomous IoT devices would require direct device-to-device communication with each other requiring different secret keys from the attribute authority (AA) based on their attributes to decrypt data. In such cases, the AA may become a bottleneck for issuing secret keys and we will need techniques to distribute secret keys in a scalable and efficient manner. 
Potential solutions for this include Hierarchical ABE (HABE)~\cite{wang2010hierarchical} and decentralizing multi-authority ABE (DMA-ABE)~\cite{lewko2011decentralizing}. In short, the HABE scheme manages the workflow in a hierarchical structure with each domain authority serving a set of domain users, whereas the DMA-ABE scheme decentralizes the single centralized AA to multiple AAs. 
Second, when an access policy needs to be updated, due to the limited storage space of IoT devices, the re-encryption of the data based on the new policy is hard to be operated locally.
A solution for this has been proposed by Huang \textit{et al.}~\cite{huang2018decent}, which designs a set of policy updating algorithms that allow the re-encryption to be operated at untrusted remote servers without breaching the privacy of the encrypted data.
The third and perhaps the greatest challenge is the issue of limited resources in IoT devices. It has been demonstrated that most classical CP-ABE schemes can hardly fit the smartphone devices and IoT devices such as Intel Edison board and Raspberry Pi~\cite{wang2014performance,ambrosin2015feasibility,ambrosin2016feasibility}. To solve this, the most common approach is to outsource the most consuming operations of ABE to powerful nodes in the network~\cite{oualha2016lightweight,yang2017lightweight}. In case that such powerful nodes are not available, Yao \textit{et al.}~\cite{yao2015lightweight} proposed a lightweight no-pairing ECC-based ABE scheme to reduce the power consumption.

\noindent \textbf{Lightweight cryptographic candidates}: As can be seen, most traditional cryptographic primitives are not applicable over resource-constrained devices. Hence, IoT privacy creates a critical need for lightweight cryptographic solutions. A non-exhaustive list of lightweight cryptographic candidates can be found in~\cite{CryptoLUX}. The design of lightweight block ciphers, based on the classification in~\cite{CryptoLUX}, consists of the Substitution-Permutation Networks (SPN) family and Feistel Networks family. 
The SPN-based schemes usually apply the S-boxes and P-boxes to perform confusion and diffusion respectively and can be roughly divided into three categories, namely the AES-like schemes (e.g., KLEIN~\cite{Gong2011}), schemes with Bit-Sliced S-Boxes (e.g., PRIDE~\cite{Albrecht2014}) and other schemes (e.g., PRESENT~\cite{Bogdanov2007}). The schemes based on the Feistel Networks split the input block into two sides, permute one with the other and then swap them. They can be designed to only use modular Addition, Rotation and XOR (e.g., RC5~\cite{Rivest1994}) or not (e.g., DESLX~\cite{Leander2007}). These lightweight schemes usually apply smaller block sizes lower than 128 bits as AES or simpler rounds without S-boxes or with smaller S-boxes to reduce the resource requirements~\cite{McKay2016}.
The lightweight hash functions are designed based on either the Merkle-Damg\r{a}rd or P-Sponge and T-Sponge. The existing lightweight hash functions such as PHOTON~\cite{Guo2011} and SPONGENT~\cite{Bogdanov2011} have already been able to achieve under 2000 GE with $0.18 \mu m$ technology for 128 digest size. 
In terms of lightweight stream ciphers, the Grain~\cite{Hell2008}, MICKEY~\cite{Babbage2008} and Trivium~\cite{De-Canniere} have stood out since 2008. In addition, recently, the NIST published its report on lightweight cryptography~\cite{McKay2016} and recommended the General Design Considerations for the future design:

\begin{mdframed}[innerleftmargin=1.6pt]
\begin{packed_enum}
  \item  \textbf{Security strength}: The security strength should be at least 112 bits.
  \item  \textbf{Flexibility}: Algorithms should be executable over an assortment of platforms and should be configurable on a single platform.
  \item  \textbf{Low overhead for multiple functions}: Multiple functions (such as encryption and decryption) should share the same logic.
  \item  \textbf{Ciphertext expansion}: The size of the ciphertext should not be significantly longer than the plaintext.
  \item  \textbf{Side channel and fault attacks}: Algorithms should be resilient to the side channel and fault attacks.
  \item  \textbf{Limits on the number of plaintext-ciphertext pairs}: The number of plaintext/ciphertext pairs processed should be limited by an upper bound.
  \item  \textbf{Related-key attacks}: Algorithms should be resilient to the related-key attacks, where the relationship between multiple unknown keys is used by the adversary. 
\end{packed_enum}
\end{mdframed}

\noindent \textbf{Middleware-blinding PETs in IoT}: The homomorphic encryption~\cite{Gentry2009}, as the most fundamental building block of the Middleware-blinding PETs, is a suite of cryptographic techniques that enable the decrypted results of computations over the ciphertext to match the results of computation over the plaintext. Its characteristics make it the best solution for outsourcing private data to untrusted parties to get their service without compromising privacy, which refers to blinding the middleware in IoT domain. Homomorphic encryption was proposed as early as 1978 but it was not until the year 2009 that the first plausible solution of the fully homomorphic encryption was proposed by Craig Gentry~\cite{Gentry2009}. Unlike the partially homomorphic cryptosystems such as the ones based on Paillier cryptosystem~\cite{Paillier1999} that support a small number of operations, the fully homomorphic encryption can enable both addition and multiplication operations over ciphertexts and therefore arbitrary computations. However, although the efficiency of the fully homomorphic encryption has been significantly improved, it is still too time-consuming for most applications. Therefore, in many cases, the partially homomorphic cryptosystems are still the preferred solution. 
IoT can benefit a lot from the homomorphic encryption~\cite{Oleshchuk2009} as well as the secure multi-party computation (SMC) schemes in the context of service discovery, data retrieval, data sharing and data outsourcing. Although most of the applications interact closely with the middleware layer, the encryption of private data is usually implemented in the perception layer. In~\cite{Malina2016}, the Paillier's partially homomorphic scheme was tested and the results showed that the scheme is still heavy for the resource-constrained devices.

\section{Privacy at Networking Layer}
\label{section5}
In this section, we discuss the secure communication and anonymous communication in the networking layer that support the \textit{Hide} and \textit{Minimize} strategies respectively.

\subsection{Secure communication}

\begin{figure}
\centering
{
   
    \includegraphics[width=3cm,height=4cm]{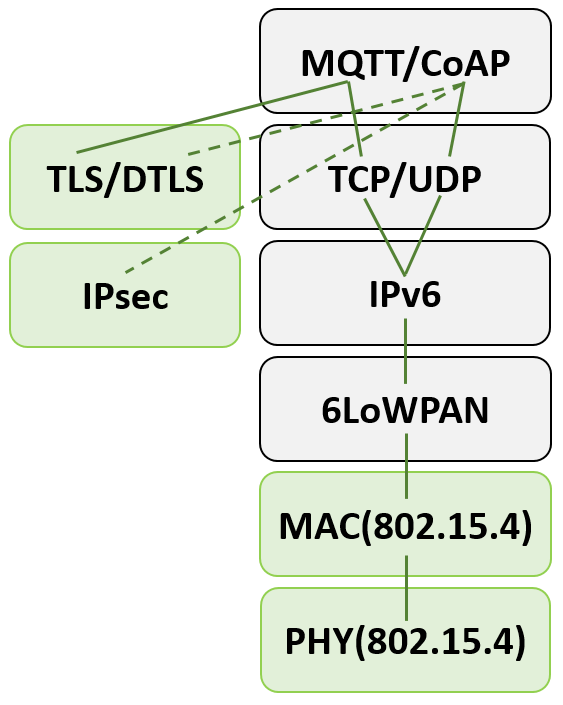}
}
\caption {Reference IoT protocol stack~\cite{Seo2005,6LoWPAN,Dierks1999,ishaq2013ietf}} 
\label{s4} 
\end{figure}

In the traditional Internet with TCP/IP stack, the communication is usually secured by either IPsec~\cite{Seo2005} in the network layer or TLS~\cite{Dierks1999} in the transport layer. In the context of IoT, due to numerous devices with constrained power, the protocol stack has to be adapted to support the transmission of IPv6 over IEEE 802.15.4 PHY and MAC, which is enabled by the adoption of 6LoWPAN~\cite{6LoWPAN} as an adaptation layer between them. A reference IoT protocol stack is shown in Fig.~\ref{s4}, which is mainly based on the IETF LLN protocol stack~\cite{ishaq2013ietf}. Above the network layer, TCP and UDP in the transport layer support different IoT application layer protocols, such as Message Queue Telemetry Transport (MQTT)~\cite{Banks2014} and Constrained Application Protocol (CoAP)~\cite{CoAP}, respectively. In terms of security, as pointed out by RFC 4944~\cite{RFC-4944} and other literature~\cite{Raza2010,Raza2014}, the AES-based security modes provided by the IEEE 802.15.4 that can support confidentiality, data authenticity and integrity, have some shortcomings. That is, the IEEE 802.15.4 only provides hop-by-hop security that requires all nodes in the path to be trusted without host authentication and key management. It may be acceptable for isolated WSNs, but not for the Internet-integrated WSNs when the messages have to travel over an IP network. Therefore, security mechanisms are required to be implemented in the higher layers to provide end-to-end security. Like the traditional Internet, the potential options include the IPsec in the network layer and the TLS/DTLS in the transport layer, where TLS and Datagram TLS (DTLS)~\cite{Rescorla2012} support TCP and UDP, respectively. The TLS/DTLS solution is the default security option of most common IoT application protocols. For example, the MQTT Version 3.1.1~\cite{Banks2014} claimed that it should be the implementer's responsibility to handle security issues and then recommended the TLS and registered TSP port 8883 for MQTT TLS communication. In contrast, the CoAP is secured by DTLS as it transmits messages over the unreliable but simpler UDP~\cite{CoAP}. The various security modes allow the devices to have either a list of pre-shared symmetric keys or a pair of asymmetric keys with or without an X.509 certificate. In addition to DTLS, the CoRE working group also proposed a draft for using CoAP with IPsec~\cite{Bormann2012}. The adoption of IPsec can make use of the built-in link-layer encryption hardware and perform transparently towards the application layer. However, due to its well-known issues with using the firewalls and Network Address Translation (NAT), the IPsec is not always available. In addition, the configuration and management of IPsec in IoT is very difficult due to the huge number of heterogeneous devices~\cite{alghamdi2013security}.

Based on the IETF protocol stack, there are some other IoT protocol stacks proposed by other standardization bodies and industry alliances. We briefly review some representatives among them.
The Thread stack~\cite{Thread} adopts 6LoWPAN to support IPv6 and leverages DTLS to secure UDP. The Thread stack has been widely adopted for connecting home devices and applications.
The IPSO Alliance~\cite{ipso-alliance} argued that using standardized protocols (e.g., IETF stack) may fail to ensure interoperability at the application layer. They proposed the IPSO Smart Objects, an object model that provides high-level interoperability between applications and devices. The core idea is to leverage the open Mobile Alliance Lightweight Specification (OMA LWM2M) on top of CoAP to enable device management operations such as bootstrapping and firmware updates. Again, DTLS is in charge of security.
The Industrial Internet of Things (IIoT) was proposed by the Industrial Internet Consortium (IIC), with the aim to connect industrial objects to enterprise systems and business processes~\cite{IIOT}. Its reference architecture adopts DDSI-RTPS~\cite{DDSI}/CoAP for UDP and MQTT/HTTP for TCP, respectively. Therefore, its security requires both TLS and DTLS.

\subsection{Anonymous communication}
The end-to-end security provided by either IPsec or TLS/DTLS can only hide the content of the messages, but not the meta-data, such as the identity (e.g., IP) of the two sides or the time, frequency and amount of the communications. Therefore, PETs enabling anonymous communication are required to handle the privacy problem due to the disclosure of meta-data, especially the identity of the initiator of the communication. For example, when health data or smart home data has to be sent to the middleware layer to get some service, it is better to make the data subject anonymous so that the personal health condition or living habits cannot be easily linked to the data subject. Such an objective can be achieved through the implementation of the anonymization and perturbation mechanisms in the perception layer, but the anonymous communication makes it also possible to handle in the networking layer.

\begin{figure}
\centering
{
   
    \includegraphics[width=8cm,height=4cm]{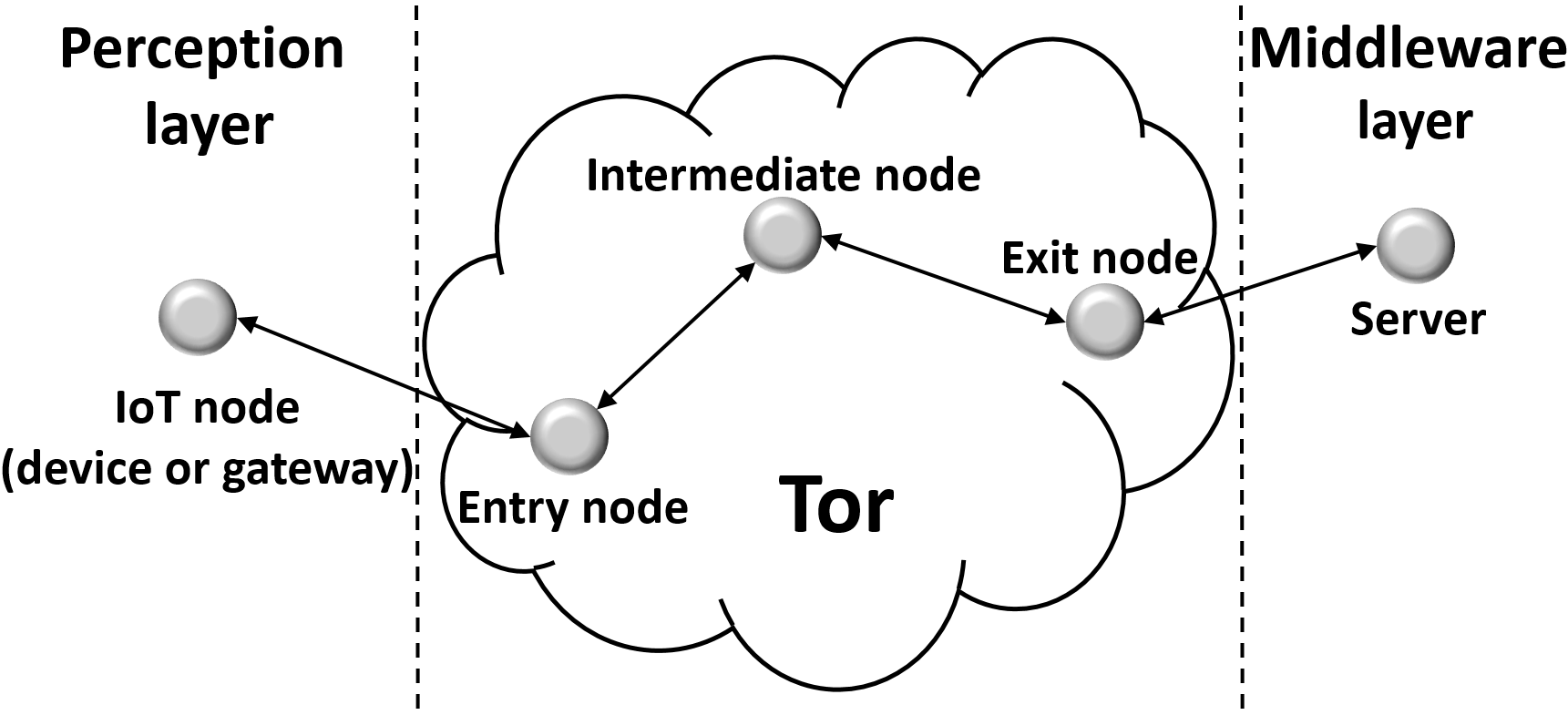}
}
\caption {Tor over IoT~\cite{ENISA2015-2,Hoang2014}} 
\label{s4-2} 
\end{figure}

The communication can be anonymized through the Proxy, the Virtual Private Network (VPN) and the onion router (Tor)~\cite{ENISA2015-2,Hoang2014}. Among them, Tor is considered an important anonymous communication PET because of its strong attack resilience~\cite{Hoang2015}. 
We show a potential Tor-based anonymous communication framework in Fig.~\ref{s4-2}. An IoT node, either a device or a gateway, wants to communicate with the middleware to get service without revealing its identity (e.g., IP address). For this purpose, instead of directly communicating with the middleware, the IoT node can first connect with the Tor network to anonymize itself. The Tor network is a distributed network with thousands of volunteers all around the world performing as the onion routers~\cite{Dingledine2004}. Its scale, as monitored by the torstatus website, is around 7000-8000 nodes in 2018~\cite{TorState}. To process the request of the IoT node, Tor will build a path (circuit) formed by one entry node, one or multiple intermediate nodes and one exit node. The raw package sent by the IoT node is then encrypted by the public keys of the nodes on the path one by one, from the entry node to the exit node, forming a layered structure, just like an onion. Each node on the path, on receiving a package from its predecessor, should decrypt one layer of the package with its private key, learn the IP of its successor and transmit the decrypted package to the successor. Each node on the path only knows the IP of its predecessor and successor and hence, the IP address of the IoT node is only revealed to the entry node and the middleware only knows the IP address of the exit node. 

The implementation of Tor over smart home was evaluated in~\cite{Hoang2015} in which, Tail, a subproject of Tor, was set up to be the central smart home gateway passed by all the outgoing data packages generated by the appliances. The results showed that Tor works well for multimedia transmission (smart TV) but not the voice-over-Internet protocol application such as Skype, due to the short time-to-live duration of UDP packets. This work demonstrated the practicability of Tor in IoT. However, several key challenges still need to be addressed. First, the access point to the Tor network should be designed to make it available to the capacity-constrained IoT devices. Second, as Tor does not support UDP, for the devices unable to encapsulate the UDP into TCP packets, mechanisms are required to enable UDP transmission over Tor. Third, the affordability of the Tor network in terms of the massive data generated by the billions of IoT nodes should be evaluated.

\begin{figure}
\centering
{
   
    \includegraphics[width=8cm,height=3cm]{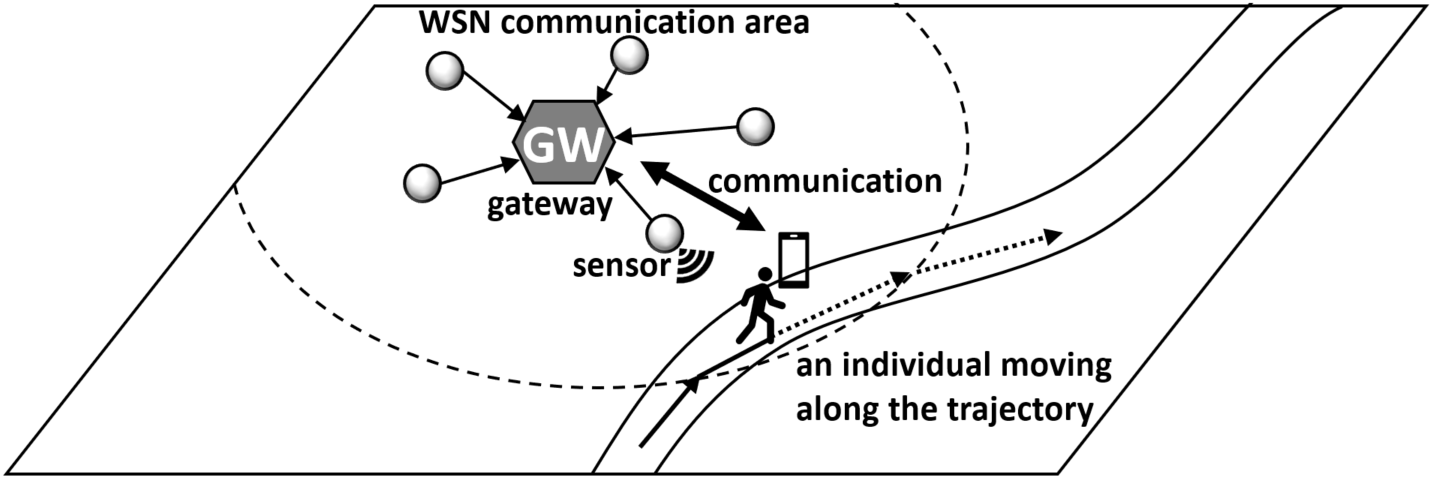}
}
\caption {Notification in the WSN} 
\label{s4-3} 
\end{figure}

\section{Privacy at Middleware Layer}
\label{section6}

In this section, we present the interaction-enhancing PETs fulfilling \textit{Inform} and \textit{Control} strategies and discuss the compliance-enhancing PETs enabling \textit{Enforce} and \textit{Demonstrate} strategies. We evaluate existing middlewares on their support for these four process-oriented strategies.

\subsection{Interaction-enhancing techniques}
\label{section41}

The main objective of interaction-enhancing techniques is to break the isolation between data subjects and their data so that data subjects can track the status of their data (\textit{Inform} strategy) and also remotely control their data (\textit{Control} strategy). 
The GDPR~\cite{GDPR} requires data subjects to get notification both before and after the data collection. Before the data collection, in addition to the data collection notification itself, data subjects should also be notified more information such as identity and contact details of data collector and purpose of the processing (Article 13). After the data collection, \textit{Inform} strategy can be combined with \textit{Control} strategy to assist data subjects to safeguard their rights, such as the right of access (Article 14), right to rectification or erasure of personal data and restriction of processing (Article 15) and right to know the personal data breach (Article 30). 

In the traditional Internet, \textit{Inform} strategy is easy to be implemented because it is the data subjects who actively determine whether to click the link to enter a website. The PETs such as the P3P~\cite{Cranor2002b} aim to assist the end users with little privacy knowledge or with no patience to quickly understand the privacy condition of the visiting websites in an automatic and usable manner~\cite{Spiekermann2009}.
Specifically, the privacy policies provided by most websites are both long and obscure with dense legalese, which makes the visitors hard to understand how their private data such as browsing history is handled. The P3P solved this problem by providing both a computer-readable format for websites to standardize the privacy policies and a protocol for the web browsers to understand the privacy policies and automatically process them based on the pre-determined privacy preference.
Unfortunately, things become harder in IoT. Unlike the traditional Internet where the end users can easily interact with the websites through static web browsers, it is essential to figure out how to effectively build the communication between data subjects and data controllers in dynamic IoT scenarios to enable \textit{Inform} and \textit{Control} strategies. To build such a communication for active collection is not hard. An example is the privacy coach~\cite{Broenink2010}, a phone application to help end users decide whether to buy products with RFID tags by actively reading RFID tags to learn corresponding privacy policies. However, to do the same thing for passive collection is more challenging. Consider the example in Fig.~\ref{s4-3} where an individual quickly passes a WSN area, the gateway has to quickly and actively get connected with the personal phone to notify the data collection, get the consent and leave information for future notifications. All these should be completed within a short period of time before the communication is disconnected. 

\begin{figure}
\centering
{
   
    \includegraphics[width=5.5cm,height=3cm]{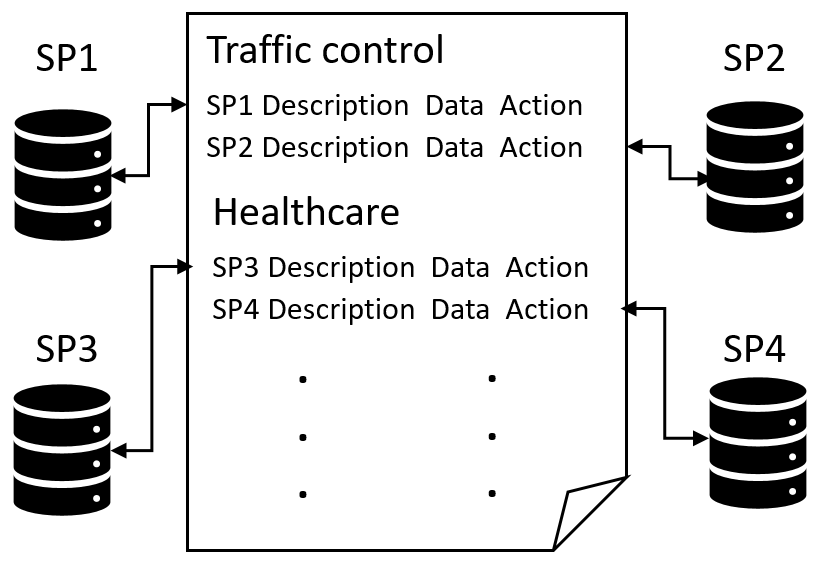}
}
\caption {Central control platform} 
\label{s4-4} 
\end{figure}

For \textit{Control} strategy, the main challenge is not how to technically implement the actions such as revision and deletion but how to design a centralized platform to simplify the control of data subjects when there are multiple data controllers. 
In active collection, each data subject can actively upload private data for different data controllers to a common personal space in cloud to simplify the tracking and control of their data~\cite{Desruelle2012,Henze2016}. In passive collection, as personal data of a data subject may be passively uploaded by data controllers to different storage places, a centralized user control platform is required, such as the one in Fig.~\ref{s4-4}. A data subject, after login, should be able to check the list of his/her personal data collected by different data controllers. Each data controller, after collecting the data, should report the collection to this central platform, link its database to the platform and provide APIs to allow the authorized data subjects to control their data. The format of a report should contain identity of the data collector, description of collection purpose, collected data and a list of possible actions that can be made by data subjects. Then, data subjects can remotely revise or delete their data.

\subsection{Compliance-enhancing techniques}

The goal of compliance-enhancing techniques is to enforce and demonstrate compliance with the privacy policy. 
The \textit{Enforce} and \textit{Demonstrate} strategies are highly related. 
First, the \textit{Enforce} strategy requires a privacy policy compatible with laws to be in place and a set of proper PETs to technically enforce it in engineering so that a data controller has the ability to comply with privacy laws. 
We require the \textit{Demonstrate} strategy here to enforce it so that the data controllers can technically prove their compliance.

As the first step, a privacy policy should be in place to guide the processing of private data. By considering personalization, this privacy policy can be replaced by a privacy preference in many cases to also reflect personal privacy demands. Such a privacy preference should be in place during the entire lifecycle of the personal data~\cite{Berghe2006}. That is, even if the personal data is disseminated from the initial data controller to the others, the privacy preference of the original data subject should be simultaneously transmitted along with the data. In other words, the privacy preference should be stuck to the corresponding data in the complicated middleware layer, which can be supported by the PET. Such a scheme was named sticky policy~\cite{Mont2003}.
The privacy model of sticky policy requires data to be first encrypted by data subjects. Then, the encrypted data and the sticky policy are sent to the data controller while the decryption key is sent to a Trusted Third Party (TTP).
Any party who wants to decrypt the data, including the initial data controller and later ones, should submit a request to the TTP with the sticky policy and credentials.
The TTP will then check the integrity and trustworthiness of them to decide whether the decrypted key can be given. During the whole process, data subjects can join or check the decision making through the TTP.
To sum up, the privacy preference must also flow along with the data and its existence should be enforced and monitored by the TTP. 
A similar approach was proposed in~\cite{Henze2016}, where the data is encrypted by the data subjects at their gateways and attached with semantic data handling annotations as the privacy preference.

After the privacy preference is in place, the PETs that can fulfill the privacy preference are required. To make it automatic, the sticky policy is recommended to be used as machine-readable semantic annotations that can be parsed by the middleware to configure the corresponding PETs. The implementation of the policy can be supported by access control mechanisms~\cite{ENISA2015}. In terms of purpose limitation, the mechanism proposed in~\cite{Mont2006} require the data requesters to declare their purpose of usage and the range of required data so that the current data controller is able to compare the declaration with the sticky annotations to make decisions. Another choice is the Hippocratic database~\cite{agrawal2002hippocratic}. As a database designed to fulfill the Fair Information Practices~\cite{Gellman2016} and especially the purpose limitation, the Hippocratic database requires the queries to be tagged with a purpose and only access the columns and tuples matching the purpose.

Finally, the most common solution to verify the compliance is the audit mechanism. That is, any interaction with private data should either be pre-checked or logged for later inspection. An example of pre-checking is the sticky policy~\cite{Mont2003}, where data requesters must first submit the sticky policy and credentials to the TTP and accept the inspection of TTP about their environment. An audit approach using the log was proposed in~\cite{Henze2016}, where personal data is encrypted in personal sphere by a gateway and then stored in a cloud platform. The cloud platform offers a database abstraction layer that can log every access of a data controller to the data with detailed information such as the access time and purpose. Next, the data subject should verify that the usage of the data complies with the privacy preference. However, even with the log information and available source code of the service, data subjects may not have the expertise to audit it. Therefore, a trusted auditor is deployed to verify the data usage in the service implementation by checking the source code. 

\subsection{Evaluation of existing middlewares}

\begin{table}[tp]  
  \centering   
  \caption{Evaluation of middlewares ($\surd$ supported with PETs, \Circle\ mentioned without details, $\times$ not mentioned)}  
  \label{tab:44} 
\begin{tabular} 
%{cccccc}  
{p{2.1cm}<{\centering} p{0.6cm}<{\centering} p{0.7cm}<{\centering} p{0.7cm}<{\centering} p{0.7cm}<{\centering} p{1.4cm}<{\centering}}
\hline
\hline 
{\bf Middleware} & {\bf year} & {\bf inform} & {{\bf control}} & { {\bf enforce}} & {\bf demonstrate}\\ 
\hline
COUGAR~\cite{Bonnet2001} & 2001 & $\times$  & $\times$  & $\times$ & $\times$\\
Impala~\cite{Liu2003} & 2003 & $\times$  & $\times$  & $\times$ & $\times$\\
IrisNet~\cite{Gibbons2003} & 2003 & $\times$  & $\times$  & \Circle & $\times$\\
Adaptive~\cite{Huebscher2004} & 2004 & $\times$  & $\times$  & $\times$ & $\times$\\
TinyLIME~\cite{Curino2005} & 2005 & $\times$  & $\times$  & $\times$ & $\times$\\
Melete~\cite{Yu2006} & 2006 & $\times$  & $\times$  & $\times$ & $\times$\\
SENSEI~\cite{Tsiatsis2010} & 2010 & $\times$  & $\times$  & $\times$ & $\times$\\
UbiROAD~\cite{Terziyan2010} & 2010 & $\times$  & $\times$  & $\times$ & $\times$\\
\hline
GSN~\cite{aberer2006middleware} & 2006 & $\times$  & $\times$  & $\times$ & $\times$\\
Xively~\cite{Xively} & 2007 & $\times$  & $\surd$  & \Circle & $\times$\\
Paraimpu~\cite{Paraimpu} & 2012 & $\times$  & $\surd$  & \Circle & $\times$\\
Webinos~\cite{Desruelle2012} & 2012 & $\surd$  & $\surd$  & $\surd$ & $\times$\\
OpenIoT~\cite{Soldatos2015} & 2013 & $\times$  & $\times$  & $\times$ & $\times$\\
Google Fit~\cite{GoogleFit} & 2014 & $\surd$  & $\times$  & $\times$ & $\times$\\
Calvin~\cite{Calvin} & 2015 & $\times$  & $\times$  & $\times$ & $\times$\\
Node-RED~\cite{Node-Red} & 2015 & $\times$  & $\times$  & $\times$ & $\times$\\
\hline
OpenHAB~\cite{OpenHAB} & 2010 & $\surd$  & $\surd$  & $\surd$ & $\surd$\\
AllJoyn~\cite{AllJoyn} & 2013 & $\Circle$  & $\Circle$  & $\Circle$ & $\Circle$\\
NOS~\cite{sicari2016security} & 2016 & $\surd$  & $\surd$  & $\surd$ & $\times$\\
\hline
\hline 
\end{tabular}
\end{table}

Currently, only a few middlewares support privacy protection. Among 61 middlewares reviewed by a recent survey~\cite{Razzaque2016}, only eight of them were labeled to support privacy. We evaluate their performance over the \textit{inform}, \textit{control}, \textit{enforce} and \textit{demonstrate} strategies. As can be seen in the first part of Table~\ref{tab:44}, among the eight middlewares, only the IrisNet mentioned the importance of the enforcement of privacy policies. In the second part of Table~\ref{tab:44}, we present middlewares reviewed by another recent survey~\cite{Ngu2017}. In Xively, permission is not required for data collection and sharing, but users are allowed to review, update or change their data in the account, which satisfies the \textit{control} strategy. Similar to Xively, the Paraimpu middleware tries to support user privacy according to the privacy laws. Both Xively and Paraimpu have the privacy policy, but the details on the enforcement are not clearly presented. The Webinos middleware can meet the three strategies in terms of protecting user privacy. In Webinos, applications require permission to access the private data. The private data is processed and stored in a local Personal Zone Proxy (PZP) and a remote Personal Zone Hub (PZH) so the users can fully control their data. Besides, through the eXtensible Access Control Markup Language (XACML) and the Webinos policy enforcement framework, users can define fine-grained access control policies that will be enforced by the PZP and PZH to mediate every access to a Webinos API.  

Additionally, we have reviewed some other IoT middlewares and software frameworks regarding their adoption of the \textit{inform}, \textit{control}, \textit{enforce} and \textit{demonstrate} strategies. The results are shown as the third part of Table~\ref{tab:44}.
The OpenHAB~\cite{OpenHAB} is a software framework designed for managing home automation systems. It makes all the devices and data stay in the local network and provides a single channel to enter the local network. It allows users to decide automation rules and has the ability to enforce the rules. It provides logging information for user-defined rules. Therefore, it satisfies all the four strategies.
The AllJoyn~\cite{AllJoyn} is a software framework aimed to create dynamic proximal networks by enhancing interoperability among devices and applications across manufacturers. Such proximal networks can make private data stay inside the local network and therefore has the potential to satisfy all the four strategies.
The middleware based on NetwOrked Smart objects (NOS)~\cite{sicari2016security} extracts privacy information from incoming data as part of security metadata at the Analysis layer, which is then used to annotate the data at the Data Annotation layer. It requires users to actively register and input private information to annotate their data. Further, the privacy protection can be enforced by the Integration layer and thus, the NOS-based middleware satisfies the three strategies.

In summary, we found that not all middlewares emphasize privacy protection. Although the recent middlewares have better protection than the previous ones, there are still privacy requirements that may be implemented at the middleware layer through PbD privacy strategies.

\section{Privacy at Application Layer}
\label{s7}

The unprecedented proximity between physical and digital worlds facilitated by IoT creates a huge number of applications~\cite{Atzori2010,Weinberg2015}. Different IoT applications may face different kinds of privacy risks as data collected in IoT applications may contain sensitive information related to the users.
For instance, in smart home applications, religious beliefs of users may be inferred from smart refrigerators and similarly, daily schedules of users may be inferred from smart lamps. 
In automobile driving applications, dozens of internal sensors monitor data related to vehicle speed and seatbelt usage that can be used by insurance companies to determine insurance premium for the users.
In healthcare and fitness applications, wearable devices may collect data that may reflect users' health information~\cite{meingast2006security}. 
Similarly in smart meters, by applying energy disaggregation over the power usage data, it may be possible to learn when and how a home appliance was used by the residents~\cite{ukil2014iot}.
In general, many of the application-level privacy risks can be handled at lower layers of the IoT architecture stack using PETs presented in Section~\ref{section4} to Section~\ref{section6}.
For example, software frameworks such as OpenHAB~\cite{OpenHAB} can make smart home a personal sphere so that data can be securely stored locally and any interaction with the data can be examined and logged.
As another example, differential privacy mechanisms~\cite{dwork,McSherry2007} can be applied to perturb the smart meter data~\cite{sankar2013smart,zhao2014achieving}, where the injected noises can be added by an in-home device.
However, it is important to ensure that the PETs employed to achieve the privacy goals does not adversely affect the utility of the target IoT application. For example, perturbation PETs such as differential privacy when applied to healthcare data that require high accuracy to be retained, the resulting perturbed data may not retain the desirable clinical efficacy and as a result, it may lead to lower application utility~\cite{fredrikson2014privacy}. In such cases, a cross-layer understanding of the impact of the employed PETs on the application-level utility is critical in determining the privacy-utility tradeoffs while designing the applications.

\section{Related work}
\label{s8}
Research on privacy in IoT has become an important topic in the recent years. A number of surveys have summarized various challenges and potential solutions for privacy in IoT. %In this section, we briefly review these surveys.
Roman \textit{et al.}~\cite{roman2013features} analyzed the features and challenges of security and privacy in distributed Internet of Things. The authors mentioned that data management and privacy can get immediate benefit from distributed IoTs as every entity in distributed IoTs has more control over the data it generates and processes. 
In ~\cite{weber2010internet}, the authors discussed several types of PETs and focused on building a heterogeneous and differentiated legal framework that can handle the features of IoT including globality, verticality, ubiquity and technicity.  
Fink \textit{et al.}~\cite{fink2015security} reviewed the challenges of privacy in IoT from both technical and legal standpoints. Ziegeldorf \textit{et al.}~\cite{Ziegeldorf2014} discussed the threats and challenges of privacy in IoT by first introducing the privacy definitions, reference models and legislation and reviewed the evolution of techniques and features for IoT.
In both~\cite{yang2017survey} and~\cite{mendez2017internet}, security risks, challenges and promising techniques were presented in a layered IoT architecture but the discussion on privacy protection is limited to the techniques related to security problems.

Although most of the existing surveys review privacy in IoT from either a technical standpoint or a legal standpoint, to the best of our knowledge, none of the existing surveys analyzed the IoT privacy problem through a systematic fine-grained analysis of the privacy principles and techniques implemented at different layers of the IoT architecture stack. In this paper, we study the privacy protection problem in IoT through a comprehensive review of the state-of-the-art by jointly considering three key dimensions, namely the state-of-the-art principles of privacy laws, architecture of the IoT system and representative privacy enhancing technologies (PETs). Our work differentiates itself by its unique analysis of how legal principles can be supported through a careful implementation of various privacy enhancing technologies (PETs) at various layers of a layered IoT architecture model to meet the privacy requirements of the individuals interacting with the IoT systems.

\section{Conclusion}
\label{s9}
The fast proliferation of low-cost smart sensing devices and the widespread deployment of high-speed wireless networks have resulted in the rapid emergence of the Internet-of-things. 
In this paper, we study the privacy protection problem in IoT through a comprehensive review of the state-of-the-art by jointly considering three key dimensions, namely the architecture of the IoT system, state-of-the-art principles of privacy laws and representative privacy enhancing technologies (PETs). We analyze, evaluate and compare various PETs that can be deployed at different layers of a layered IoT architecture to meet the privacy requirements of the individuals interacting with the IoT systems.
Our analysis has shown that while many existing PETs (e.g., differential privacy, Tor) demonstrate a great potential for use in the IoT, the adoption of these techniques requires a careful consideration of the unique features associated with the IoT, including the use of heterogeneous power-limited devices and the massive need for streaming data flow. 
We expect this study to provide a broader understanding of the state-of-the-art principles in privacy legislation associated with the design of relevant privacy enhancing technologies (PETs) and how privacy legislation maps to privacy principles which in turn drives the design of necessary privacy enhancing technologies to be employed in the IoT architecture stack.

\renewcommand\refname{Reference}
\footnotesize
\bibliographystyle{IEEEtran}
\urlstyle{same}
\bibliography{main.bib}

\vskip 0pt plus -1fil

\begin{IEEEbiography} [{\includegraphics[width=1in,height=1.25in]{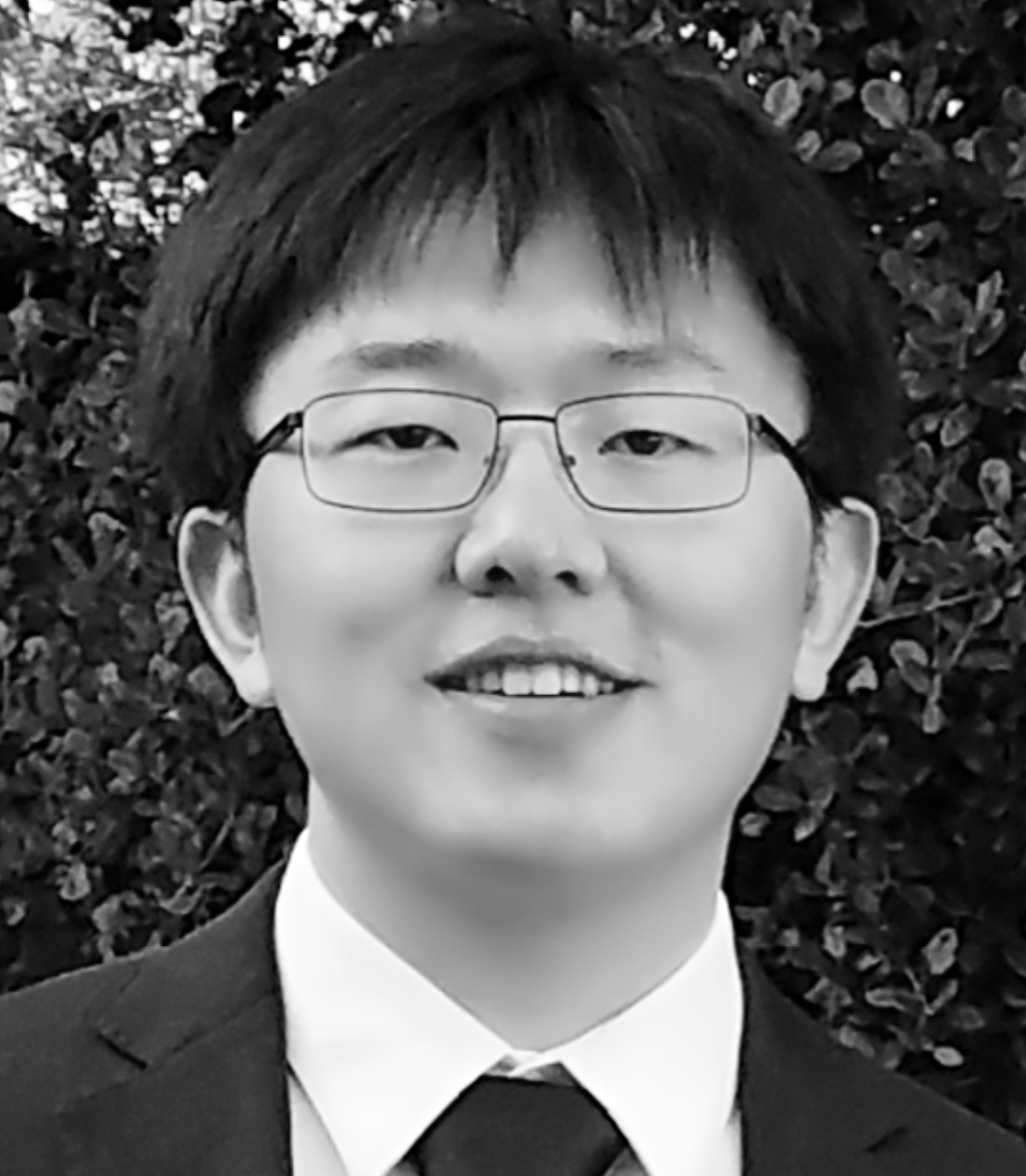}}]{Chao Li}
    is currently a 4th year Ph.D. student in the School of computing and information, University of Pittsburgh. He is also a member of the Laboratory of Education and Research on Security Assured Information Systems (LERSAIS). Before that, he got his MSc degree in Communication and Signal Processing from Imperial College London and BEng degree in Electronics and Electrical Engineering from both University of Edinburgh and Dalian University of Technology. His current research interests include location and data privacy and blockchain-based protocol design. He is a student member of the IEEE.
\end{IEEEbiography}

\vskip 0pt plus -1fil

\begin{IEEEbiography}
    [{\includegraphics[width=1in,height=1.25in]{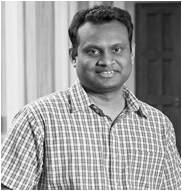}}]{Balaji Palanisamy}
 is an Assistant Professor in the School of computing and information in University of Pittsburgh. He received his M.S and Ph.D. degrees in Computer Science from the college of Computing at Georgia Tech in 2009 and 2013, respectively. His primary research interests lie in scalable and privacy-conscious resource management for large-scale Distributed and Mobile Systems. At University of Pittsburgh, he codirects research in the Laboratory of Research and Education on Security Assured Information Systems (LERSAIS). He is a member of the IEEE. Dr. Palanisamy is currently serving as an Associate Editor for the IEEE Transactions on Services Computing journal.
\end{IEEEbiography}

\end{document}